# Long-Term Stability of Graphene/c-Si Schottky-Junction Solar Cells

*Djordje Jovanović*[*,1], *Miloš Petrović*[2], *Tijana Tomašević-Ilić*[1], *Aleksandar Matković*[3], *Matevž Bokalič*[4], *Marko Spasenović*[5], *Konstantinos Rogdakis*[2,6], *Emmannuel Kymakis*[2,6], *Dragan Knežević*[7], *Lucio Cinà*[8], *and Radoš Gajić*[1]

[1]Institute of Physics Belgrade, University of Belgrade, 11000 Belgrade, Serbia
[2]Institute of Emerging Technologies (i-EMERGE) of HMU Research Center, Heraklion 71410, Crete, Greece
[3]Institute of Physics, Montanuniversität Leoben, Franz Josef Strasse 18, 8700 Leoben, Austria
[4]Faculty of Electrical Engineering, University of Ljubljana, Trzaska cesta 25, SI-1000 Ljubljana, Slovenia
[5]Center for Microelectronic Technologies, Institute of Chemistry, Technology and Metallurgy, University of Belgrade, 11000 Belgrade, Serbia
[6]Department of Electrical & Computer Engineering, Hellenic Mediterranean University (HMU), Heraklion 71410, Crete, Greece
[7]Vojnotehnički Institut, 11000 Belgrade, Serbia
[8]Cicci Research s.r.l., Via Giordania 227, 58100 Grosseto (GR), Italy

E-mail: (djordje@ipb.ac.rs)

**Abstract**

A long operational lifetime is required for the use of solar cells in real-life photovoltaic applications. The optimization of operational lifetimes is achieved through understanding the inherent degradation phenomena in solar cells. In this study, graphene/Si Schottky-junction solar cells were produced, utilizing liquid-phase-exfoliated graphene as an active surface. The operational and interface stability of these solar cells over a period of 5 years in ambient conditions (following ISOS-D protocols: dark storage/shelf life) was examined, and the origin of their degradation was reported. It was found that the dominant degradation mechanism could be attributed to the degradation of silver contacts. This was indicated by a decrease in shunt resistance, an increase in the ideality factor (due to a higher carrier recombination), and a constant defect density in graphene films for up to 4 years. Measurements across the solar cell's active area during the 5-year period revealed neither significant spatial inhomogeneity, nor shunt channel defects.

## 1. Introduction

As global energy demands increase annually in conjunction with technological advancements and a rising trend in connectivity, there is a clear need for compact and reliable clean energy sources [1]. Presently, solar energy harvesting competes favorably with other energy sources worldwide and no longer requires subsidies to remain competitive [2]. This represents a significant de-globalization trend in the energy field, driven by countries seeking solutions for energy resilience. Photovoltaic (PV) technologies have already made considerable commercial progress and are projected to contribute more than 5% of the global energy demand by 2025 [3]. Scaling up to 5 TW of solar energy production (~15% of global demand) by 2030, or even achieving 30% of the world's energy demand (~10 TW), in accordance with the EU Strategic Energy Technology Plan target by 2040–2050 [3-5], appears to be attainable. To accomplish these goals, investments in new solar/coal facilities should currently follow a ratio of 9:1 [6]. This serves as a compelling argument for additional research and development on emerging PV absorbers to diversify the existing portfolio of solar technologies [7]. Inorganic crystalline silicon (c-Si)-based single-junction PV cells currently dominate terrestrial solar energy conversion due to their high efficiency (>20%), low cost (<0.5 $/W), and good reliability (approximately 25 years) [7,8]. However, organic devices such as organic solar cells (OSC or OPV) [9], dye-sensitized solar cells (DSSC) [10], and perovskite solar cells (PSC) [11] need to achieve improved stability, environmentally safe materials, and production processes to become technologically viable [12-14].

In addition to power conversion efficiency (PCE), there are other crucial aspects to consider in PV energy harvesting, such as stability, manufacturing cost, and ease of processing [15,16]. When it comes to commercial applications, the long-term stability of solar cell (SC) devices in relation to air, humidity, temperature, and light exposure over an extended period is of utmost importance. Real-life applications necessitate long operational lifetimes for solar cell devices, and understanding and mitigating degradation phenomena are prerequisites for the successful implementation of promising technologies beyond silicon [12].

Additionally, the development of small-scale, lightweight, and portable PV devices for indoor usage, known as iPV (indoor Photovoltaic) [17,18], presents an attractive direction for meeting the energy demands of low-power consumption devices. The future market for self-powered electronics as part of the Internet of Things (IoT), including distributed sensors, remote actuators, and communication devices, holds significant potential for iPV applications [19,20]. When integrated into solar cells, graphene serves multiple functions owing to its high transparency and charge mobility, making it a viable alternative to conventional transparent electrodes such as indium tin oxide (ITO) and fluorine doped tin oxide (FTO) [21]. Moreover, a built-in electric field is developed at the graphene/silicon (Gr/Si) interface, aiding in the collection of photogenerated charge carriers [22].

Groundbreaking research on graphene/silicon (Gr/Si) junctions [23] demonstrates the integration of graphene into a Schottky junction with n-doped Si, achieving a power conversion efficiency (PCE) of 1.5%. Substantial improvements exceeding 2% have been observed under low-illumination conditions (0.15 Sun). However, a major obstacle that hampers the performance of graphene/Si solar cells is the poor fill factor [23]. This indicates that charge transport and recombination rates occur on a similar timescale, leading to competitive processes often visually represented by the appearance of an "s-kink" in the current–voltage curves [24]. To address this challenge, one approach is chemical doping, such as using trifluoromethanesulfonic acid (TFSA), which has yielded a PCE of 8.6% [25]. Another strategy involves the inclusion of a buffer oxide layer to reduce the transport barrier, resulting in a PCE surpassing 10% [24]. A more complex method for enhancing PCE involves fine-tuning the graphene Fermi level. This can be accomplished by employing graphene as a gating electrode, covering the dielectric barrier over the semiconductor substrate, leading to a high open-circuit voltage (>0.9 V) and a remarkably high PCE (18.5%) [26].

The study of graphene/Si Schottky junctions for indoor light harvesting has been limited [23], but there are reports focusing on stability. These reports indicate an operational stability of 90%, 80%, and 70% for 7, 30, and 90 days, respectively [14]. On the other hand, the performance and stability of solar cell architectures, other than graphene, in indoor applications [19,20] or stability field [9-11,27,28] have been extensively studied. Indoor photovoltaics (iPV) based on a-Si:H demonstrate a suppression of the fill factor (FF) from 56% to 53%, which then remains constant for 7 days, achieving up to 21% efficiency [29]. Commercial a-Si:H iPV under white LED or fluorescent light (FL) lighting ranges from 4.4% to 9.2% efficiency. Dye-sensitized solar cells (DSSCs) have achieved 51.6% operational stability (efficiency) after 273 hours of continuous one-sunlight soaking in hybrid devices [10]. In iPV, DSSCs exhibit 34% efficiency for 1000 lx FL, with extrapolated lifetimes of 25–40 years for DSSCs with hydrophobic dyes. Organic photovoltaics (OPVs) have demonstrated 1,330 hours of operational stability at the maximum power point (MPPT) [30], and 31% power conversion efficiency (PCE) for 1650 lx white LED (WLED). Quantum dot solar cells (QDSCs) have achieved 9.7% efficiency under 1000 lx LED light for Si QD-based hybrid PVs and 19.5% efficiency under 2000 lux FL for colloidal metal chalcogenide quantum dots (QDs). Perovskite solar cells (PSCs) maintain 90% of initial efficiency over a 2400-hour test in ambient conditions with 25% humidity [27]. Encapsulated PSCs have demonstrated 18.7% PCE with 95% operational stability after 3 months of outdoor exposure [28]. Lead–halide perovskite (MAPbI3) has achieved an iPV PCE of 37.2% under 1000 lx WLED.

Our study aims to validate the long-term environmental operational stability of these devices over a 5-year period and identify the critical structural mechanisms responsible for performance degradation during the aging process. Additionally, a proposed low-cost fabrication method for graphene-based electrodes is presented, highlighting its potential for stable graphene/Si Schottky-junction solar cells. Finally, suggestions are made on mitigating these negative effects by adjusting the trade-off between device design and fabrication cost, particularly through the use of solution-based graphene fabrication. The prospective applications of such devices were found to be suitable for the compact off-grid (without battery) supply of IoT sensors in indoor conditions.

## 2. Experimental Section

### 2.1. Material, Production, and Chemicals

2.1.1. Graphene film synthesis and fabrication

Stock graphene dispersions were produced from commercially available graphite powder (Sigma Aldrich) with an initial concentration of 18 mg/ml by liquid phase exfoliation in N-methyl-2-pyrrolidone (NMP, Sigma Aldrich, product no. 328634), following the same procedure as previously reported [31-33]. Dispersions of liquid-phase-exfoliated (LPE) graphene were used for thin film formation by Langmuir-Blodgett self-assembly (LBSA) at the toluene-water interface [34]. Quartz, Si/SiO$_2$, and prepatterned Si/SiO$_2$ wafers were used as substrates. Surface modification of the prepared films was realised by annealing and UV/O$_3$ treatment. The annealing was carried out in a tube furnace at 300 $^0$C in an argon atmosphere for 2 h. UV/ozone treatment was performed by exposing the graphene films to ultraviolet radiation and ozone for 5 min at 50 $^0$C chamber temperature in a Novascan UV/Ozone Cleaner.

2.1.2. Silicon wafers

We used Si/SiO2 <100> c-Si wafers (525 μm thick) with an SiO$_2$ (300nm thick) layer on top. The Si wafers are doped with As to a concentration of $N_d$=1.2-7.4e$^{19}$ cm$^{-3}$. They have R=0.001-0.01 Ω-cm, electron affinity Xsi = 4.05 eV, electron mobility = ~ 45 cm$^2$/Vs, Kr effective Richardson constant = 120 A/cm$^2$K$^2$, relative permittivity ε = 11.7 [35] -11.9 [23], ε$_0$ = 8.854 × 10$^{-12}$ F cm−1, dEgap ~75meV, and an optical band gap Eg,1 = 1.03 [36] with a work function = 4.4 eV. The data are provided by the wafer manufacturer (Inseto, UK).

2.2. Solar cell fabrication, functionalization, and storage

SCs are formed by Schottky junctions between graphene films and highly n-doped c-Si, as schematically shown in Fig. 1a. Graphene films are deposited on pre-patterned Si/SiO$_2$ wafers (with different Si-junction (active) square areas ranging from 0.8 mm$^2$ to 1.2 mm$^2$). Contacts of the front (graphene anode) and back (Si cathode) electrodes were made with a silver paste. In order to increase device performance, the obtained graphene films were thermally annealed (A films) and then exposed to photochemical oxidation (AO films) [31]. Two different types of solar cells were assembled using A and AO graphene films. The annealing process is utilized to remove residual solvents and increase connections between the graphene layers, while the UV/ozone treatment decreased the density of the defects [31]. Solar cells were kept for 5 years on the shelf, i.e., in environmental conditions, according to ISOS-D protocols: dark storage/shelf life [37].

2.3. Material analysis

The optical and electrical characterization of A and AO graphene films was performed with UV-VIS spectrophotometry, Raman spectroscopy, and two-point probe resistance measurements. An additional analysis of pristine graphene films was performed with atomic force microscopy (AFM) and Kelvin Probe Force Microscopy (KPFM).

Optical transmittance was obtained with a UV-VIS spectrophotometer (Perkin-Elmer Lambda 4B UV-VIS) on quartz wafers. Raman was performed with a TriVista 557 S&I GmbH Micro Raman spectrometer (λ=532 nm) at room temperature, with a Coherent laser power of 20mW, using a 50X objective. The two-point resistance of each graphene film, R$_{sheet}$, was calculated by considering sample geometry factors.

The surface morphology of the films was characterized with a Horiba/AIST-NT Omegascope atomic force microscope (AFM) in tapping mode. Aseylec probes were employed (spring constant ~ 42 N/m, resonant frequency ~ 70 kHz, tip radius below 30 nm). Amplitude-modulated (AM)-KPFM measurements were carried out in a two-pass mode under ambient conditions, with the probe lifted 25 nm during the second pass. Topography and contact potential difference (CPD) images were processed in the open-source software, Gwyddion v2.56. In the case of topography images, zero-order line-filtering was applied, as well as leveling of the base plane. In the case of CPD images, only zero-order line-filtering was applied and the data were summed in a histogram. For KPFM measurements, the graphene films were grounded and supported by $SiO_2$. Before and after the work function (WF) of each graphene film was measured, a freshly cleaved HOPG surface (scanned within 5 min from cleaving) was probed with an identical set of parameters [38]. The HOPG measurements were used to calibrate the probe and confirm that the WF of the probe did not change during the measurements of the graphene films.

2.4. Device Characterization and Measurements

I–V measurements in dark and light conditions were performed with a modular testing platform (Arkeo – Cicci research s.r.l.) composed of a white LED array (4200K) with optical power density tuneable from 10 to 100 mW/cm$^2$ (0.1-1 Sun), and a high-speed source meter unit (600k samples/s) in a four-wire configuration under ambient conditions. The LED intensity was calibrated at the equivalent of 1 Sun intensity by adjusting the $J_{sc}$ value to be equal to the measured in the J–V curve using the solar simulator.

Small-perturbation transient photovoltage (TPV) measurements were conducted using the transient module of the commercial system Arkeo (Cicci Research s.r.l.). The open circuit mode was controlled via differential voltage amplifiers and short photogeneration pulses were generated using a fast LED (470 nm) with a Lambertian radiation pattern and 120° viewing angle limited at 100 mA current. The background bias was set close to the open circuit voltage ($V_{oc}$) in order to avoid the resistor–capacitor (RC) effect.

The external quantum efficiency (EQE) measurements were conducted using an integrated system (Enlitech, Taiwan), with a lock-in amplifier and a current preamplifier under short-circuit conditions. The light spectrum was calibrated using a monocrystalline photodetector of a known spectral response.

Thermal images were taken with a thermovision camera, SC 7200 FLIR, with a spectral range from 1.5 μm to 5.1 μm, an instantaneous field of view (IFOV) of 0.6 mrad, a focal plane array of 320x240, an objective of 50 mm, and with a distance to sample of 1 m.

Light Beam Induced Current (LBIC) measurements were carried out using an in-house developed LBIC system. Laser diodes with a 642 and 1060 nm wavelength were used. The response was measured using the lock-in technique at excitation frequency of 1023 Hz. The measurement step was 25 μm and 10 μm for AO and A sample types, respectively. The beam spot was estimated to a few measurement steps maximum, since there were no sharp features in the samples.

## 3. Results and discussion

According to UV-VIS spectroscopic measurements presented in Fig. 1b., A and AO graphene films both exhibited high optical transparency T>70% at a wavelength of 660 nm with constant sheet resistance $R_{sheet}$ (~10kΩ/sq). It should be noted that the measured optical absorbance is due to graphene only, because the measurements are performed on graphene supported by transparent quartz wafers. Despite the absorbance of the film itself, the graphene does not affect the Si absorption. The multilayer graphene structure was confirmed with Raman spectroscopy (Fig. 1c). The spectra feature all pronounced bands characteristic of the LPE graphene films [31]. An additional analysis of graphene films, obtained with AFM and KPFM techniques, is presented in Fig. S1 of Supporting Information (SI).

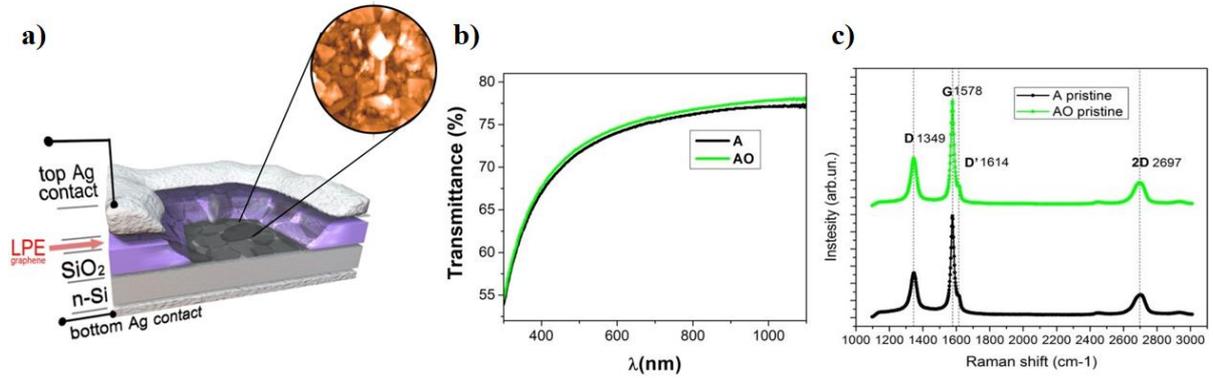

Fig. 1.a) Graphene/c-Si Schottky-junction solar cell with an AFM topography image in the inset (inset diameter 1 μm, z scale 100 nm), b) optical transmittance of graphene films on quartz substrates, c) Raman spectra of pristine A and AO graphene films on the Si/SiO$_2$ substrate.

In order to obtain the complete set of operational solar cell parameters, we performed dark I–V and transient photovoltage (TPV) measurements on the same samples at the time of fabrication, pristine (t=0) and after 1.5 years (dark storage/shelf life conditions). For the calculation of the SC parameters, we used standardized different methods [39], with the most important parameters being the series resistance ($R_s$), shunt resistance ($R_{sh}$), and ideality factor (n, used to identify the dominant form of recombination). The dark I-V measurements of pristine and 1.5 year old A and AO samples are presented in Fig. 2. $R_{sh}$ was obtained from the inverse slope of the dark I-V curve in the linear regime at V~0, when the current is at its maximum. $R_s$ was obtained from the part of the dark I-V curve with a high slope on the semi-log scale which corresponds to Fermi level alignment that occurs with adding electrons during voltage increase. Mathematically, this ratio is defined as $dV/d(\ln I) = R_s \cdot I + n \cdot (k_B \cdot T)/q$, where n is the diode ideality factor, q is elementary charge, T is absolute temperature, and $k_B$ is Boltzmann's constant. A plot of $d(V)/d(\ln I)$ vs. I will give $R_s$ as the slope and n as the y-axis intercept. n was obtained from the slope $q/(nk_BT)$, i.e., a linear fit of the [ln(I)-V] curve (Fig. 2a and b), which has a nearly linear part in the range of 0.1–0.4 V of the forward bias regime. As evident from the semi-log IV curves (Fig. 2a) and 2b)), the SC leakage current increased over time, observed from the rise in the negative section of the curve. This behavior indicates that $R_{sh}$ decreases with aging because of the creation of unwanted pathways for the current leakage [40], and it would be expected to decrease more with further cell aging. As seen in Table 1, $R_{sh}$ increases in pristine AO over A cells and decreases more rapidly in aged cells (double in A and 10 times in AO cells). Furthermore, we can see that our SC has high values of $R_{sh}$, which could be very important for indoor (low-light) performance [41-44]. In order to validate the results for $R_s$ and obtain the Schottky barrier height ($\Phi_b$), the voltage drop can be related to the deviation from an ideal diode as the Schottky barrier function: $H(I)=V-n(k_B \cdot T)/q \cdot \ln(I/(A \cdot K_r \cdot T^2))=R_s \cdot I+n \cdot \Phi_b$, where $K_r$ is Richardson's constant for n-Si [120 A/cm$^2$K$^2$], A is the active area of the cell, and $\Phi_b$ is the Schottky barrier height [34]. Using the values of n obtained in previous $dV/d\ln(I)$ vs. I analysis, the y-intercept gives $\Phi_b$ and the slope validates $R_s$. On the other hand, $R_s$, obtained via two complementary methods, is maintained almost constant with values ~10 kΩ, with a little drop over time, more so in doped AO cells (Fig. 2c) and d)). The $R_s$ value of ~10 kΩ of solar cells is exclusively connected to the high resistance of graphene films ($R_{sheet}$ ~10 kΩ/sq). Constant $R_s$ over time is associated with stable charge transport across interfaces within the PV cell. The constant $R_s$ is a good indicator that there was no interface degradation in time between the Si, graphene, and silver (Ag) contacts. $R_s$ was smaller in AO than in A cells, both for pristine and aged cells, pointing toward better electrical properties of the AO cells. It should be noted that a small divergence in $R_s$ between the cells can also arise from the fabrication of the cell (incomplete removal of native oxide), and intrinsic defects close to the surface of the Si wafer. Small divergences are not always related to the graphene $R_{sheet}$ being constant in both samples; hence, an insufficiently clean Si surface may leave an inhomogeneous native $SiO_2$ layer on the Si surface, or there may occur incomplete removal of the wafer cleaner, i.e., acetone. It must be pointed out that the obtained values for $R_s$ are three orders of magnitude higher than in other graphene/Si solar cells made with high-quality CVD graphene [23-26]. The high $R_s$ could be the consequence of the cost-effective solution-based LPE process that results in films made of interconnected small flakes of graphene, rather than a clean homogeneous film such as CVD graphene. From Fig. 2, we can also observe that the ideality factor decreases in AO relative to the A pristine solar cells (1.89 to 1.37), which are values that are in agreement with literature reports for non-treated A cells (range of 1.6−2.0) and doped AO cells (range of 1.3−1.5) [24]. Over time, n increases in both cells, but more so in AO-treated cells. As the ideality factor is very sensitive on surface recombination [45,46], we can hypothesize that the level of recombination could be affecting the change in the ideality factor.

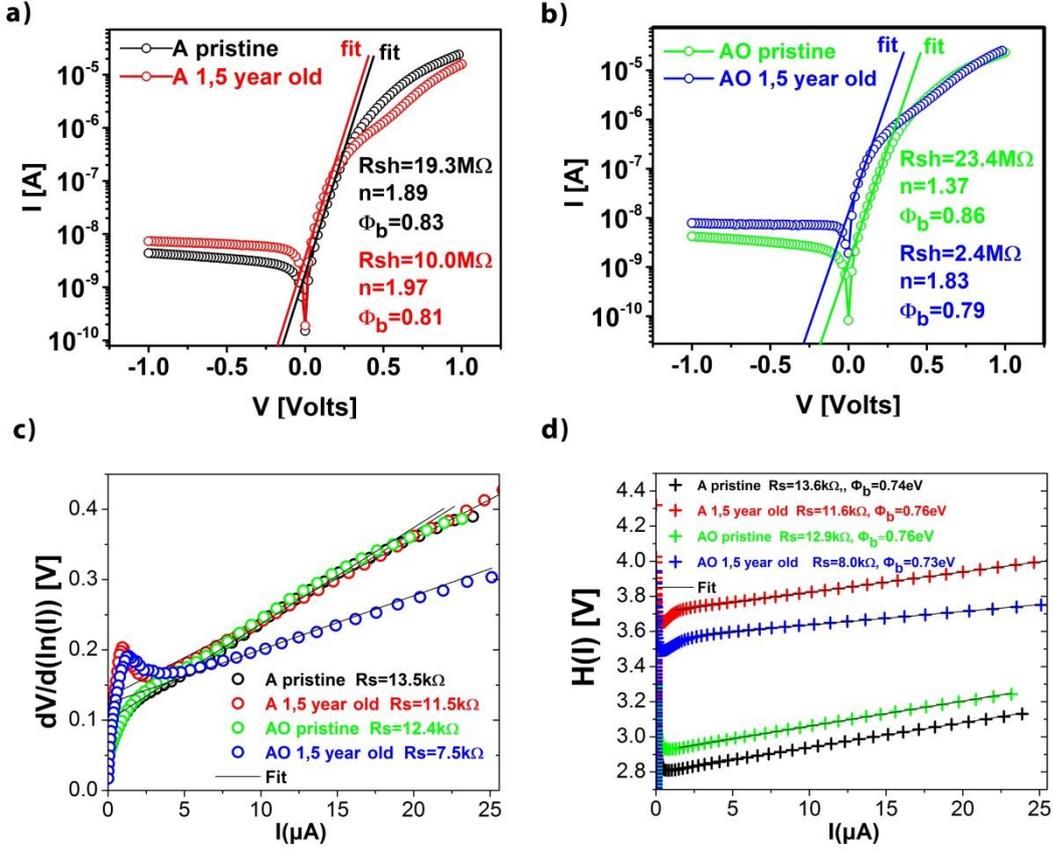

Fig. 2. Dark I–V measurements for pristine and 1.5-year-old A and AO solar cells. The fitted values of $R_{sh}$ and $\Phi_b$ are extrapolated from lnI(V) for a) A and b) AO cells. c) The fitted values of $R_s$ are extrapolated from dV/dlnI(I), and d) validated values of $R_s$ and $\Phi_b$ are extrapolated from H(I).

The assumption of stable charge transport across interfaces within the PV cell, based on the stable $R_s$ in time, is further supported by the obtained values of $\Phi_b$ (see Table 1 and Fig. 2a), b), and d)). The $\Phi_b$ is obtained from the saturated current ($I_s$) as the y-intercept of the linear part of the semi-log I–V curve according to the formula $\Phi_b=(k_B \cdot T)/q \cdot \ln((A \cdot K_r \cdot T^2)/I_s)$. From Fig. 2 and Table 1, we can see that $\Phi_b$ remains almost the same in both diodes and in time (0.79-0.86eV). This indicates that aging did not impact the materials at the interfaces [40]. $\Phi_b$ is estimated to be 0.83 and 0.86 for pristine, and 0.81 and 0.79 for aged cells. The small rise in $\Phi_b$ with the UV/ozone for pristine samples could be associated with hole-doping [25,32]. A decreasing trend for $\Phi_b$ is observed in both cells, but more so in AO-treated cells as a consequence of a higher degradation of AO graphene films. Contrary to this, if we use values for the graphene work function, $W_{gr}$, obtained from KPFM (4.98 and 4.81 for A and AO films, Fig. S1), and based on the Schottky–Mott model, where $\Phi_b = W_{gr} - \chi_{Si}$ ($\chi_{Si} = 4.05$, electron affinity of Si), we have the opposite trend for $\Phi_b$. $\Phi_b$ decreases by ozone-doping from 0.93 to 0.76 for AO and A pristine samples, respectively [23,25]. Due to electron-doping of graphene films in environmental conditions and the work function in metals, in practice, exposure to ozone does not have a real impact on $\Phi_b$. The reason for this is that the Schottky theory neglects the impact of the different surface states on $\Phi_b$ (additional interfaces formed because of incomplete covalent bonds, existence of impurities, patina, and basically everything which generates additional charges on phases boundaries) [47]. Constant $\Phi_b$ in time does not mean that the material did not degrade. As we mentioned earlier, it could be from the $SiO_2$ or remaining impurities that result in a relatively high $\Phi_b$. Aging degradation does not have enough influence to perturb the $\Phi_b$ value. Furthermore, $\Phi_b$ can be unstable, varying during an applied voltage scan as a consequence of the varying Fermi level. A high-defect population or existence of impurities can impact carrier accumulation. This carrier accumulation can occur on the Si surface because of the forming of $SiO_2$ or a high mobility difference between the Si and graphene. On the other hand, since $\Phi_b$ and the built-in potential $V_{bi}$ are connected by the formula $\Phi_b = V_{bi} + e^{-1} k_B T \ln(N_c/N_d)$ [48,49], where $N_c$ is the effective density of the states in the conduction band and $N_d$ is the doping level of the semiconductor, the increase in $\Phi_b$ could be a consequence of the increase in $V_{oc}$ and $V_{bi}$. This $\Phi_b$ increase leads to an increased charge transfer across the metal–semiconductor interface, creating a larger potential drop $V_{bi}$ across the depletion region, and allowing for a more efficient collection of electrons and holes [25]. This requires additional attention, since the difference between the metal–semiconductor Fermi level does not mean that is the real value for $V_{bi}$. In highly doped materials, as in our case ($N_d \sim 10^{19} cm^{-3}$), electrons and holes can tunnel through this layer, decreasing the total potential below the theoretical expected value. Furthermore, the structure of graphene has a big impact on $\Phi_b$, since defects can ruin the ideality factor [50].

Table 1: Dark measurement parameters of pristine t=0 and aged t=1.5 year old A and AO cells obtained with different theoretical methods.

| Parameter (Method) | Cell | | | |
|---|---|---|---|---|
| | A | | AO | |
| Year | pristine | 1.5 | pristine | 1.5 |
| $R_s$ (dV/dlnI)[kΩ] | 13.50 | 11.54 | 12.39 | 7.46 |
| $R_s$ (H(I) validate) [kΩ] | 13.61 | 11.58 | 12.89 | 8.05 |
| $R_{sh}$ [MΩ] | 19.3 | 10.0 | 23.4 | 2.4 |
| n(lnI(V)) | 1.89 | 1.97 | 1.37 | 1.83 |
| $\Phi_b$ (lnI(V)) [eV] | 0.83 | 0.81 | 0.86 | 0.79 |
| $\Phi_b$ (H(I) validate) [eV] | 0.74 | 0.76 | 0.76 | 0.73 |

In order to validate the conclusion that the ideality factor increases as a consequence of more intense recombination, TPV measurements were implemented. Small-perturbation measurements were used to probe the bimolecular charge recombination rate and estimate the carrier lifetime. In general, a longer carrier lifetime means that the recombination order is reduced (recombination speed is inversely proportional to the lifetime) [25]. This leads to higher-quality cells in operational terms. In Fig. 3, the results for the lifetime of both sample types and aging periods are presented. We observe a drop in lifetime (about 50%) between the pristine and aged devices in both UV/ozone and non-treated devices. These results are in agreement with the I–V dark measurements ($R_{sh}$ and n), where, in time, the recombination increases as a consequence of the degradation of material (most probably of silver contacts). The drop in lifetime is smaller between A and AO cells (~30%), and lower in A cells that have a longer lifetime than AO cells. A little shorter AO lifetime is related to the oxygen treatment of graphene [31], where oxygen is introduced as a recombination centre. As mentioned earlier, the carrier lifetime is linked with the direct non-radiative recombination of carriers, which means that in the case of aged cells, we have suppressed carrier transport between the electrodes and a more intense recombination. According to the experimental conditions, the shallow defects do not have an impact on this because their population is filled before the TPV measurements. It could be expected that the carrier lifetime decreases further as the cells are aged even more.

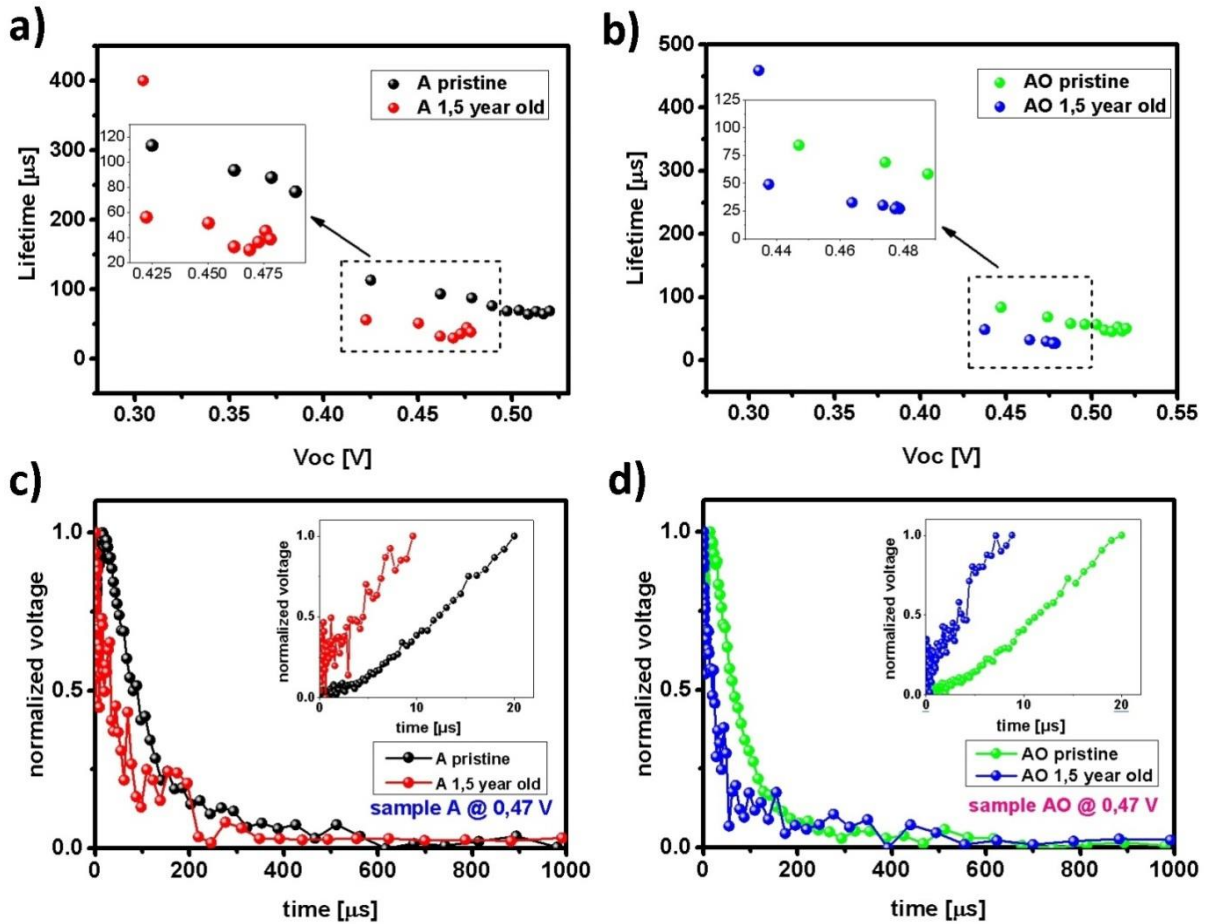

Fig. 3. TPV measurements for A and AO pristine and 1.5-year-old samples. a) and b) lifetime as a function of $V_{oc}$; c) and d) photovoltage as a function of time. Insets-speeds of voltage signal.

The insets of Fig. 3c) and d) present the spread of the voltage signal that increases in real conditions, potentially originating from the low electrode contact quality (silver paste). In both cases, the charge carrier lifetime is about 20 μs for pristine and 10 μs for aged samples. The smaller slope in aged cells confirms a small variation in $R_s$ obtained from dark current analysis. This means that pristine devices have more surface defects and the aging process accelerates some kind of ``passivation process`` that enables the decreasing in defects. The effect is most likely caused by either the carrier's reservoir generation or the space charge region at the interface.

With I-V dark and TPV solar cell measurements on pristine and 1.5-year-aged solar cells, we can conclude that there was material degradation without an interface degradation between the graphene and Si ($R_s$, $\Phi_b$ ~ const., Table 1). We assume that the primary material degradation bottleneck could come from the degradation of silver contacts ($R_{sh}$ decreases, ideality factor increases (higher carrier recombination), and the slopes of voltage are higher over time (Fig. 2 and 3)).

By utilizing highly n-doped c-Si wafers, a higher concentration of carriers can be achieved. This effect becomes more prominent in dark conditions, where photogeneration phenomena are absent. The increased carrier concentration enhances mobility. However, at higher intensities, when the concentration of electrons/holes rises, recombination also increases, leading to a rapid decrease in power conversion efficiency (PCE), primarily due to a decrease in the fill factor (FF). The FF represents the balance between electrode transport and lifetime. Fig. S2 illustrates that mobility exponentially declines as $N_d$ (carrier concentration) increases, impacting recombination as a result of an increased number of free electrons. Essentially, there exists a trade-off between these two processes. Consequently, it can be suggested that for indoor low-light applications such as IoT devices, employing highly doped Si wafers with high-resistivity ($R_s$) graphene films could be a suitable option for solar cells.

The tested devices, which were stored in accordance with ISOS-D protocols (dark storage/shelf life [37]), demonstrated operational stability for a minimum of 4 years. The stability of the graphene films was assessed using Raman spectroscopy, specifically by examining the ratio of the peak intensities D/G and D/D' (calculated from the peak maximum intensity), as shown in Fig.4. For the pristine samples, we observed a consistent value for defect density (D/G ~0.4) and the D/D' ratio, which was approximately 4. The value of the D/D' ratio is indicative of the predominant defect type in the graphene samples [51,52]. The D/D' ratio suggests that edges are the primary defect type, aligning with our previous findings for the thin films derived from liquid-phase-exfoliated graphene [31]. However, after 4 years, there was an increase in both the D/G and D/D' ratios, indicating the onset of film degradation [31,47,48]. Further film degradation may occur with time, which leads to an estimated lifetime of these solar cells on the order of 10 years.

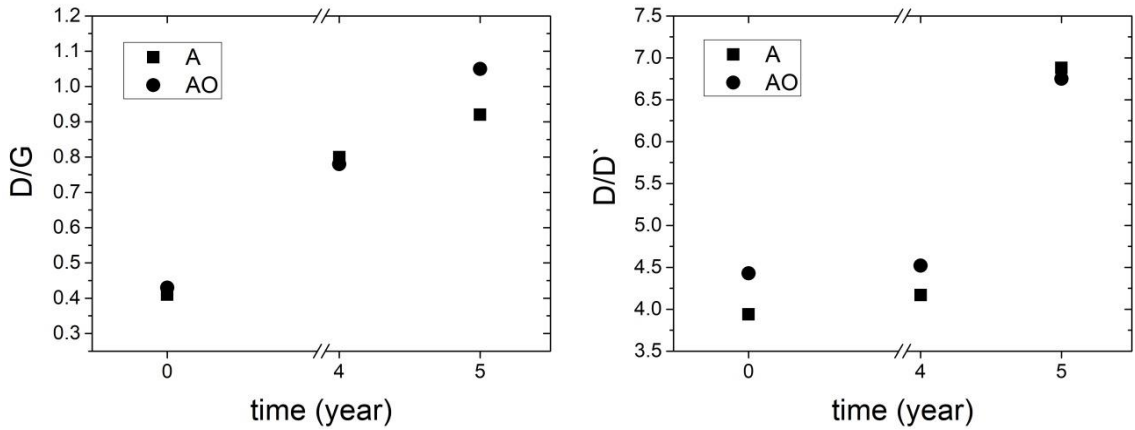

Fig. 4. D/G and D/D` ratios over time for A and AO graphene films.

Additionally, the spatial homogeneity of the 4-year-old devices was assessed using Light Beam Induced Current (LBIC) measurements. The results, depicted in Fig. 5, illustrate the LBIC scans of two devices of different types (top: A–thermally annealed, bottom: AO–thermally annealed and UV/ozone-treated) conducted at two laser wavelengths (left: 642 nm, right: 1060 nm). Notable inhomogeneities are not observed within the active area of the devices, indicating a relative spatial stability. However, minor inhomogeneities are present in the 1060 nm scan of device A, which are absent in the 642 nm scan. These discrepancies may originate from the inhomogeneities on the back side of the device. Since LBIC scans were not conducted in the beginning of experiment, it is not possible to determine their temporal origin.

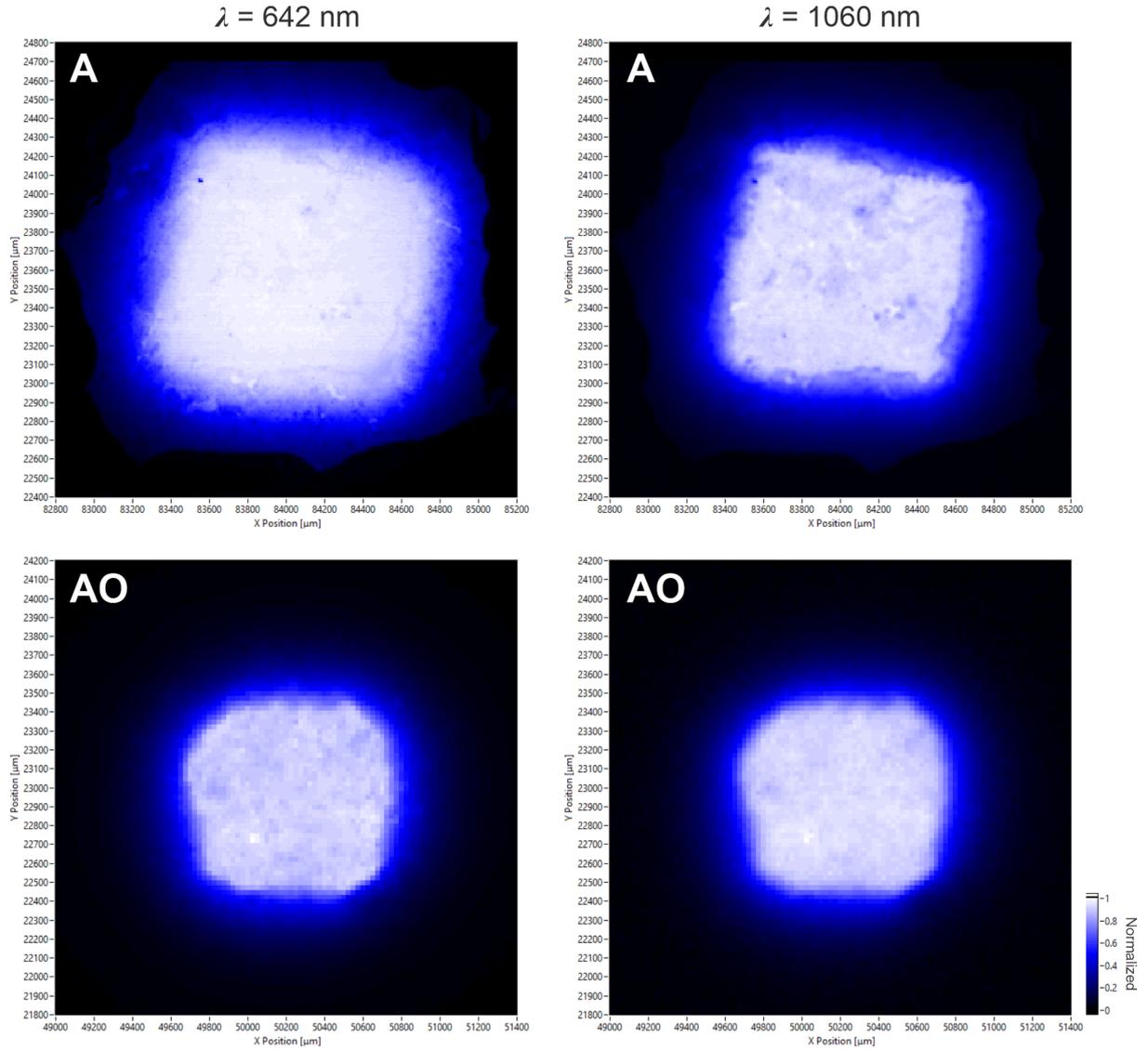

Fig. 5. LBIC spatial maps of 4-year-old A and AO solar cells, measured at 642 and 1060 nm laser wavelength.

An additional analysis of the stability of solar cells aged for 3 and 5 years was performed using thermovision (further details in Supplementary Information), as shown in Fig. S3. Furthermore, the effect of ageing is explored by measuring of external quantum efficiency (EQE) in pristine and 5 years old A and AO cells (Fig. S4). We can see that EQE decreased about 50% in both cells (more in A then AO cells) as the effect of ageing of graphene films and silver contacts (see Fig. 4 and Fig. S5). Since the operation of our novel solar cells in bright conditions is not the main subject of this paper, we have measured the photovoltaic response only on the two cells discussed above.

Finally, the current–voltage response of both the fresh and aged devices (pristine and 1.5 year old) under variable light bias (0.13–1.0 Sun, white LED 4200K source) is presented in Fig. S5 in Supplementary Information. It is evident that the aged cells exhibit about 50% degradation in efficiency and JV compared to the fresh cells. Additionally, we observe that the efficiency increases as solar intensities decrease, particularly in indoor conditions. This phenomenon can be attributed to factors such as a high $R_{sh}$, lower recombination intensities in solar cells, and the high $N_d$ of the Si wafers [41-43].

## 4. Conclusion

In summary, we have validated the operational stability of graphene/Si Schottky-junction solar cells over 5 years, and graphene/Si interface stability over 1.5 years in air ($R_s$ and $\Phi_b$ ~ const.), respectively. A critical structural mechanism responsible for the performance degradation during the aging process was identified to be the degradation of Ag contacts ($R_{sh}$ decreases, n increases, and the slopes of voltage are higher over time). LBIC and FLIR techniques validated the spatial stability, homogeneity, and the absence of shunt channel defect places over a 5-year period, while Raman spectroscopy confirmed the low defect density of graphene films over 4 years. The use of highly doped Si wafers with high $R_s$ graphene films could be a good option for IoT Gr/Si solar cell indoor applications. This study therefore proposes a facile, simple, and low-cost fabrication approach of graphene-based electrodes with long-term stability. Our work highlights the long-term stability of liquid-phase-exfoliated materials in an operational environment, which was observed before [34] and indicates that LPE followed by Langmuir–Blodgett assembly is an efficient fabrication method for real-world devices.

**Credit author statement**

**Dj. Jovanović**: Conceptualization, Project administration, Funding acquisition, Data curation, Formal Analysis, Investigation, Methodology, Validation, Visualization, Writing-original draft and Writing-review and editing. **M. Petrović**: Conceptualization, Data curation, Formal Analysis, Methodology, Investigation, Visualization, Writing-original draft and Writing-review and editing. **T. Tomašević-Ilić**: Formal analysis, Investigation, Visualization, Resources, Data Curation, Writing-original draft and Writing-review and editing. **A. Matković**: Conceptualization, Formal analysis, Investigation, Visualization, Resources, Data Curation, Writing-original draft and Writing-review and editing. **M. Bokalič**: Formal analysis, Investigation, Visualization, Resources, Data Curation, Writing-original draft and Writing-review and editing. **M. Spasenović**: Conceptualization, Formal analysis, Writing-original draft and Writing-review and editing. **K. Rogdakis**: Investigation, Formal analysis, Project administration, Writing-original draft and Writing-review and editing. **E. Kymakis**: Resources, Supervision, Funding acquisition, Writing - review and editing. **D. Knežević**: Formal analysis, Investigation, Visualization, Resources, Data Curation, Writing-original draft and Writing-review and editing. **L. Cinà**: Investigation, Formal analysis, Visualization, Writing - review and editing, **R.Gajić**: Supervision, Funding acquisition, Writing - review and editing.

**Acknowledgments**

The authors acknowledge the funding provided by the Ministry of Education, Science, and Technological Development of the Republic of Serbia. M.S acknowledges the funding provided by the Ministry of Education, Science, and Technological Development of the Republic of Serbia through project 451-03-47/2023-01/200026. Dj.J acknowledges the partial funding provided under COST Action MP1406 and support from a number of individuals whose efforts provided an additional value to the project, including T. Maksudov and G Kakavelakis from Hellenic Mediterranean University, Crete, Greece; Nikola Tasić from University of Belgrade and University of Ljubljana; Ivana Validžić from University of Belgrade; Nicola Lisi from ENEA, Italy; Aleksandar Radulović from AKR Konsalting, Serbia. Dj.J is grateful to God, his parents, family, and friends for long-time support during this project. This work is dedicated to his passed parents.

**Conflict of Interest**

The authors do not have any financial/commercial conflicts of interest.